\documentclass[12pt]{iopart}
\newcommand{\Kz}{K$^0_{\mathrm S}\ $}
\newcommand{\nL}{$\Lambda$\ }
\newcommand{\Jpsi}{J/$\psi$\ }
\newcommand{\npt}{$\rm p_T$ }
\newcommand{\raa}{$\rm R_{AA}$ }
\newcommand{\iaa}{$\rm I_{AA}$ }
\newcommand{\npi}{$\pi$\ }

\usepackage{graphicx}   

\begin{document}

\title{ALICE results from the first Pb--Pb run at the CERN LHC}

\author{J. Schukraft for the ALICE Collaboration}

\address{CERN Div. PH, 1211 Geneva 23}
\ead{schukraft@cern.ch}
\begin{abstract}

After 20 years of preparation, the dedicated heavy ion experiment ALICE took first data at the CERN LHC accelerator with proton collisions at the end of 2009 and with lead beams at the end of 2010. This article will give a brief overview of the main results presented at the Quark Matter 2011 conference.

\end{abstract}


\section{Introduction}
After 1987 in Nordkirchen and 2001 in Stony Brook, the 2011 conference in Annecy is the $3^{rd}$ Quark Matter where results from a new heavy ion facility are presented for the first time. After 20 years of design, construction and preparation of the experiments and the accelerator, the LHC started operating in November 2009 with pp collisions at $\sqrt{s} = 900$ GeV and reached its current maximum energy of 7 TeV in March 2010. The primary ALICE goal for pp was to collect about $10^{9}$  minimum bias (MB) collisions under clean experimental conditions.
The total pp data sample collected at 7 TeV (900 GeV) corresponds to some 800 M (8 M) MB triggers, 100 M muon triggers and about 20 M high multiplicity events. In addition, a short pp run was taken in March 2011 at the Pb--Pb energy of 2.76 TeV to collect a minimum set of comparison data (recording 70 M MB events and integrating 20 nb$^{-1}$ of luminosity for rare triggers). Early physics results from the proton runs are summarized in~\cite{Schukraft:2010ru}.

The first heavy ion run at 2.76 TeV/nucleon took place in November 2010, after only a few days of switching over from the pp set-up. As was the case for proton beams, the LHC worked exceedingly well also for heavy ions, with a steeply increasing luminosity which reached about $2~10^{25} \rm cm^{-2} \rm s^{-1}$ towards the end of the run and yielded some 30 M nuclear MB interactions on tape. Initial results on global event characteristics (charged particle multiplicities, elliptic flow, volume and lifetime inferred from pion HBT, and high \npt suppression) became available already during or shortly after the run~\cite{Schukraft:2011kc}. At this conference, a second wave of results from ALICE will be presented in close to 40 talks and 70 posters. The following will be a very brief 'guided tour' through some of the highlights, while more details, references and figures can be found in the individual ALICE contributions to these proceedings.

\subsection{The ALICE experiment}
ALICE, which stands for A Large Ion Collider Experiment, is very different in both design and purpose from the other experiments at the LHC~\cite{Schukraft:2010rt,Evans:2009zz}. It is specifically optimized to study heavy ion reactions, but in addition data-taking with pp (and later p-nucleus) is required primarily to collect comparison data for the heavy ion program~\cite{Carminati:2004fp}.

\begin{figure}[!t]

\centerline{\includegraphics[width=1.0\textwidth]{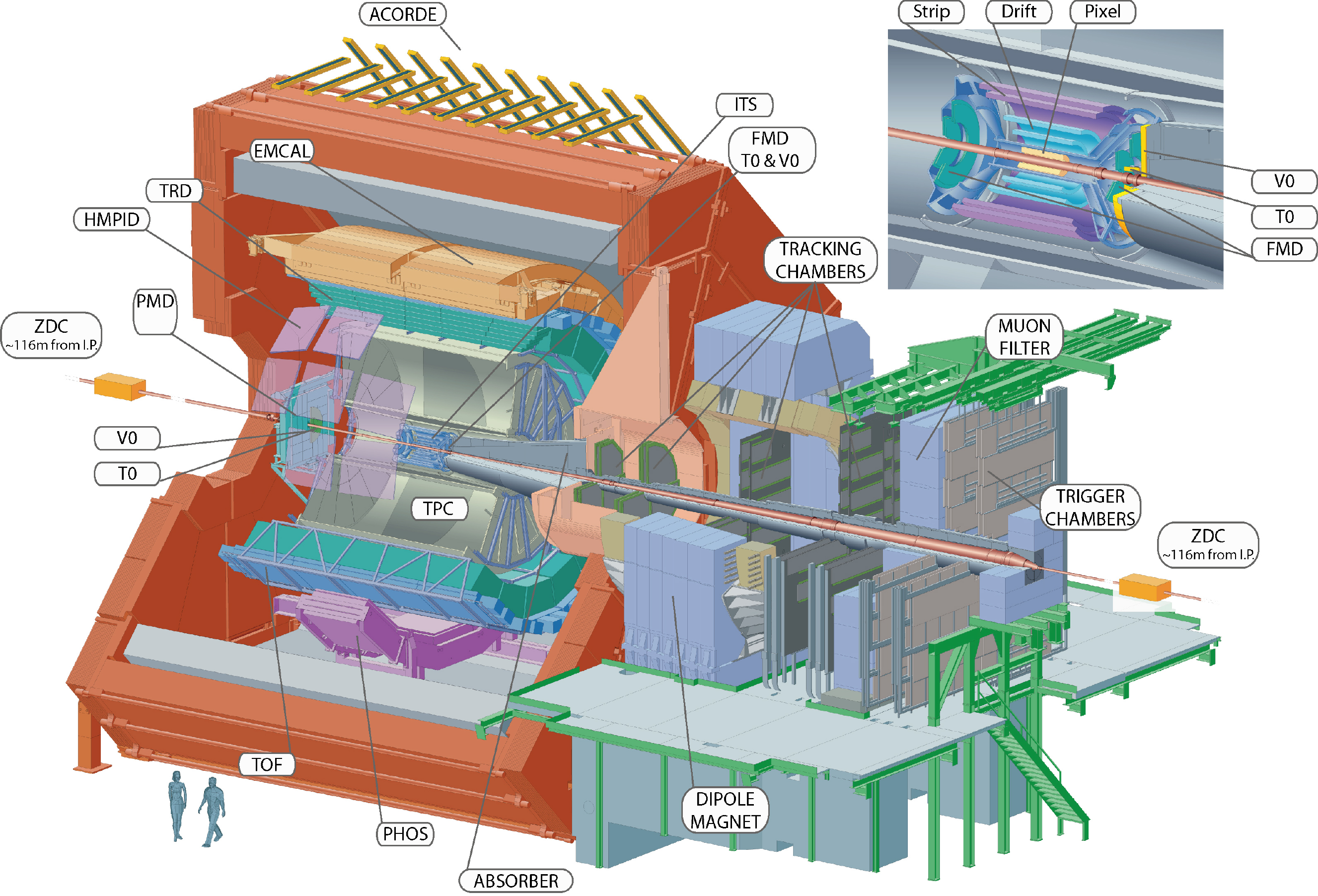}}
\caption{ALICE uses 18 different detector systems, indicated in the figure with their acronyms. The central detectors are located inside the L3 magnet (left side), the forward Muon arm spectrometer is located on the right and the insert top right is a blow-up of the interaction region showing the 6 layers of the silicon vertex detector (ITS) and the forward trigger and multiplicity detectors (T0, V0, FMD).
}
\label{setup}
\end{figure}

A schematic view of the detector is shown in Figure~\ref{setup}.
ALICE consists of a central part, which measures hadrons, electrons and photons, and a forward spectrometer to measure muons. The central part, which covers polar angles from $45^{\circ}$ to $135^{\circ}$ over the full azimuth, is embedded in the large L3 solenoidal magnet (B = 0.5 Tesla). It consists of an inner tracking system (ITS) of high-resolution silicon detectors (two layers each of pixel, drift, and double-sided strips), a cylindrical TPC, three particle identification arrays of Time-of-Flight (TOF), Cherenkov (HMPID) and Transition Radiation (TRD) counters, and two single-arm electromagnetic calorimeters (high resolution PHOS and large acceptance EMCAL). The forward muon arm ($2^{\circ}-9^{\circ}, 2.5 < \eta < 4$) consists of a complex arrangement of absorbers, a large dipole magnet (3 Tm field integral), and 14 stations of tracking and triggering chambers. Several smaller detectors for triggering and multiplicity measurements (ZDC, PMD, FMD, T0, V0) are located at small angles. 
The main design features include a robust and redundant tracking over a limited region of pseudorapidity, designed to cope with the very high particle density of nuclear collisions, a minimum of material in the sensitive tracking volume ($\approx$10\% radiation length between vertex and outer radius of the TPC) to reduce multiple scattering, and several detector systems dedicated to particle identification over a large range in momentum.

The layout of the ALICE detector and its eighteen different subsystems are described in detail in~\cite{ALICEdet}. The experiment is fully installed, commissioned and operational, with the exception of two systems (TRD and EMCAL) which were added more recently. Both systems had about 40\% of their active area installed during 2010; the EMCAL was completed during the winter shutdown 2010/11 and the TRD will be completed early next year.

\section{Overview of physics results}

\subsection{Global properties, spectra, particle ratios}
The first, and most anticipated, result from LHC concerned the charged particle multiplicity density in central collisions. The value finally measured with Pb--Pb at 2.76 TeV/nucleon, ${\rm d}N_{\rm ch}/{\rm d}\eta \approx 1600$, was somewhat on the high side of more recent (post-RHIC) predictions but well below the maximum values ($>4000$) anticipated during the design phase of ALICE in the early 90's. From the measured multiplicity in central collisions one can derive a rough estimate of the energy density, which is at least a factor 3 above RHIC. The corresponding increase of the initial temperature is about 30\%, even with the conservative assumption that the formation time $\tau_0$ does not decrease from RHIC to LHC.

When combined with lower energy data, the charged particle production per participant pair, $(dN_{ch}/d\eta)/(0.5 N_{part})$, is seen to rise with centre of mass energy $\sqrt s$ approximately like $s^{0.15}$, stronger than in pp ($s^{0.11}$). Somewhat surprising is the fact that the centrality dependence is practically identical to Au--Au at RHIC (at least for $N_{\rm part} >50$), despite the fact that impact parameter dependent shadowing/saturation could be expected to be much stronger at LHC (much smaller Feynman-x). Saturation physics gives a natural explanation for both observations, as the energy dependence is predicted to be a power law with roughly the correct exponent, whereas the centrality dependence changes only weakly (logarithmically) with the saturation scale and therefore the beam energy. Note that the different energy dependence measured for pp and AA collisions requires that the centrality dependence at RHIC and LHC must be slightly different for very peripheral reactions, where particle production (per participant) must converge towards the corresponding pp value.

The large multiplicity at LHC, together with a large detector acceptance, brings some practical benefits e.g.  for the centrality measurement (resolution reaches $<0.5\%$ for most central), the event plane determination for flow (resolution correction factor up to 0.95), or in general all fluctuation measures where resolution is typically proportional to the square root of multiplicity.

The freeze-out volume and total lifetime of the created system was measured with identical particle interferometry (HBT) for both pp and AA collisions. Compared to top RHIC energy, the 'volume of homogeneity' increases by a factor two and the system lifetime increases by more than 30\%, pretty much in line with predictions from hydrodynamics. Also a more detailed study of the out/long/side radii as a function of pair momentum shows very good agreement with hydro models, at least with those that also describe well the RHIC data. While both pp and AA radii scale with the charged particle multiplicity to some power (close to 1/3), the scaling coefficient and the offset are different leading to different radii for the same final state multiplicity (i.e. very high multiplicity pp and peripheral AA). Therefore the freeze-out is not governed solely by a universal final state particle density, as has been sometimes advocated at lower energies, but the initial geometrical size of the colliding particles plays a role as well.

Identified particle spectra have been measured by various means (dE/dx in silicon and TPC, TOF, Cherenkov radiation and topological decay reconstruction) both as a function of centrality and transverse momentum (out to 20 GeV/c for charged pions, order 10 GeV/c for \Kz and $\Lambda$, and 3 GeV/c for charged kaons and protons). The spectral slopes change significantly compared to RHIC, most dramatically for protons, indicating a much stronger radial flow. A first attempt at a blast wave fit indicates that the average radial flow velocity reaches about 2/3 of the speed of light and that the kinetic freeze-out temperature drops below 100 MeV. While the K/\npi ratio increases slightly with centrality from the corresponding value in pp, the p/\npi ratio stays at the pp value of about 0.05; a value which is smaller by a factor 1.5-2 than the one predicted by thermal models of particle production assuming the canonical chemical freeze-out conditions expected at LHC (T $\approx 165$ MeV, $\mu_B \approx 0,  \gamma_S = 1$). However, before any strong conclusion about thermal particle ratios can be drawn, more species have to be measured and the treatment of feed down corrections at LHC (and lower energies) must be clarified.

The 'baryon anomaly' (very large p/\npi and $\Lambda$/\Kz ratios) is also present at LHC. The $\Lambda$/\Kz ratio reaches a maximum of about 1.5  around 3 GeV/c in \npt (to be compared with a maximum value of 0.5 in pp at 7 TeV). Both the maximum value as well as its position in \npt  changed only modestly compared to RHIC. However, unlike at RHIC, the enhancement of roughly a factor three relative to the pp value persist out to about 6 GeV/c.

Further information on the above topics can be found in the plenary contributions on global event properties~\cite{qmtoia}, particle interferometry~\cite{qmkisiel}, and identified particles~\cite{qmfloris}, as well as the corresponding contributions to these proceedings from parallel sessions and posters.

\subsection{Flow, correlations, fluctuations}
One of the highlights of this conference, with major contributions from all experiments at both RHIC and LHC, was on various aspects of azimuthally asymmetric flow measurements. While already the very first elliptic flow measurement $v_2$ at LHC had shown that the 'QGP' is still very much the 'strongly interacting and almost perfect liquid' discovered at RHIC, and very well described by hydrodynamics, the progress since then has been rapid and most impressive.  First precision measurements are on the horizon (e.g. reducing and/or subtracting non-flow contributions, cumulants up to $\rm 8^{th}$ order, measurements in fine centrality bins, ..), higher Fourier components ($v_3, v_4, v_5$) and event-by-event flow fluctuations have been extracted from the data, and flow-like correlations are seen up to $p_T > 10$ GeV/c, well into the region presumably dominated by path length dependent jet quenching rather than hydrodynamic flow. Flow coefficients extracted from two particle correlations (avoiding the 'near-side jet' region) are fully consistent with the standard flow measurements (cumulants, event plane, ...) for momenta up to 3-4 GeV/c. At higher momenta away-side jet correlations start to become dominant leading to a breakdown of the 'factorisation' of the two-particle correlation coefficient into two single particle flow coefficients.

The hydrodynamic origin of triangular flow $v_3$ has been clearly established by measuring the characteristic mass and \npt dependent flow pattern with identified particles ($\pi$, K, p) for the first time. The $v_3$ event plane shows little or no correlation with either the first order reaction plane or the second order event plane, as expected if the origin of $v_3$ is due to initial state shape fluctuations. Triangular flow also most naturally and economically explains the unusual structures seen already at RHIC in two-particle correlations (often called 'Mach Cone' and 'Soft Long-Range Ridge') as an interference between various flow components, chiefly $v_2$ and $v_3$. Identified particle elliptic flow at LHC shows a stronger dependence on particle mass than at RHIC, with light (pions) and heavy (protons) particles pushed further apart from each other - as expected from hydro because of the increased radial flow. This measurement also seems to indicate that the 'quark number scaling' of $v_2$ observed at RHIC is most likely accidental at low momenta ($<$ 2 GeV/c); at LHC scaling with particle mass as predicted by 'particle hydro' describes the data actually better. However, at higher \npt, quark number scaling continues to hold for $v_2$ (as well as $v_3$) and therefore a coalescence type mechanism may still be needed to describe flow in the intermediate momentum region. 

Conspicuously absent from most flow discussions is the rapidity-even first Fourier component $v_1$, which should be present in the data with a strength comparable to $v_3$. It is very visible in the two particle correlation analysis, but potentially strongly contaminated by non-flow effects. However, the rapidity-odd ($v_1(\eta)=-v_1(-\eta)$) directed flow has been measured for charged particles close to midrapidity and is found to follow an approximate 'limiting fragmentation' scaling when compared to RHIC.

Transverse momentum fluctuations have been measured in pp and PbPb; they are similar in the region of overlap when plotted as a function of particle multiplicity. However, for central AA collisions, these fluctuations show an intriguing reduction in magnitude which is much more pronounced than at RHIC. Charge dependent fluctuations have been investigated via three particle correlators. Somewhat surprisingly, the same signal (in both sign and magnitude) is seen as at RHIC, where it has been interpreted as a sign of 'local parity violation' (also called Chiral Magnetic Effect).  

While this conference laid to rest Mach Cone and Soft Ridge, it also saw the birth of 'Flow tomography', where event-by-event density distributions and fluctuations in the initial state are measured via momentum correlations in final state particles. This has great potential to lead to one of the first precision measurements of a fundamental property of the QGP, the normalized shear viscosity $\eta$/S. The various Fourier flow components have a non-trivial dependence on the initial state and therefore should be able to discriminate between different geometrical models, thus removing a major source of systematic uncertainty in the extraction of $\eta$/S. 

Flow and correlation results are discussed further in~\cite{qmsnellings,qmjanfiete}.

\subsection{High \npt suppression, heavy flavour, quarkonia}
The very first measurement of high \npt suppression has shown that $\rm R_{AA}$, the properly normalized ratio of \npt spectra in Pb--Pb and pp, shows a pronounced minimum around 5-7 GeV/c which is slightly deeper than at RHIC, but then rises significantly towards 20 GeV/c. The new results extend the \npt range towards 100 GeV/c and make use of the measured pp comparison spectrum at 2.76 TeV (because of limited statistics, the reference spectrum still needs to be extrapolated above 30 GeV/c). The rise in \raa is seen to continue up to the highest $\rm p_T$, with a gradual change in slope above 30-40 GeV/c. The centrality dependence of \raa becomes less pronounced at progressively higher momenta. The suppression for identified particles (\Kz and $\Lambda$) becomes independent of particle type above 8 GeV/c but is different below, where the kaon \raa is rather close to the one for inclusive charged particles, whereas the \nL \raa is larger and reaches unity at around 2-3 GeV/c (presumably a reflection of the 'baryon anomaly' mentioned earlier). 
The yield \iaa of particles associated to a high \npt trigger particle (normalized to the yield from pp) is compatible with unity for peripheral collisions for both near-side and away-side associated particles. For central collisions, the away-side yield shows the suppression expected from in-medium energy loss (\iaa $\approx 0.5-0.6$), similar in magnitude to RHIC, whereas the near-side is slightly but significantly enhanced (\iaa $\approx 1.2$). This could indicate some modification also of the near-side jet at LHC.

Open heavy flavour production has been measured in ALICE via fully reconstructed hadronic charm decays ($\rm D^0, D^+, D^*$) and semi-leptonic decays in electrons (central barrel) and muons (forward muon spectrometer), in both pp and Pb--Pb. In all cases, the total and differential cross-sections in pp (2.76 and 7 TeV) are at the upper edge of, but fully consistent with, perturbative QCD calculations (FONLL) and  also consistent with the results from the other three LHC experiments in regions where the phase space coverage overlaps. They can therefore be used with confidence as a basis to construct the reference data for the Pb--Pb results. The suppression of prompt (i.e. B feed-down corrected) charmed mesons has been measured in the \npt range 2-12 GeV/c. Above 5-6 GeV/c, the charm meson \raa ($\approx 0.2$) is compatible within statistical and systematic errors with the \raa of charged pions, whereas it may be somewhat larger at lower momenta.
The \raa of heavy flavour electrons and inclusive muons (which contain a mix of c and b decays) show a suppression in central collisions which is consistent with each other but larger than the \raa of prompt charm. Such a feature is present in models where the energy loss for charm is larger than for beauty. 

The picture which emerges at LHC is that already hinted at at RHIC: Quenching of charm quarks is essentially as strong as that of inclusive hadrons, at least for \npt above 5-6 GeV/c.
Qualitatively, a difference between the \raa of heavy quarks and pions  is expected as the energy loss of massive quarks should be smaller than that of massless gluons (the dominant source of pions at LHC) because of colour charge and the 'dead cone' effect. A quantitative comparison with models will be required to see if the predicted size of the effect is compatible with our data.

\Jpsi production, the classical deconfinement signal, has confounded expectations and interpretations ever since the first nuclear suppression, measured with Oxygen beams at the SPS, was announced at the Quark Matter conference in 1987. The puzzling fact that \Jpsi suppression is rather similar in magnitude at SPS and RHIC may indicate that either the \Jpsi is not suppressed at all and we see only $\psi$' and $\chi_c$ suppression at both SPS and RHIC, or that \Jpsi suppression is more or less balanced by enhanced production via coalescence of two independently produced charm quarks. A resolution to this puzzle may come from measuring \Jpsi production at LHC (where coalescence effects should be stronger because of the abundant charm production) and by comparing the suppression patterns of the \Jpsi and $\Upsilon$ families.

We have measured the differential inclusive \Jpsi cross-section down to zero \npt in pp at 2.76 and 7 TeV and found it fully consistent with pQCD calculations and results from the other LHC experiments. Using the measured pp cross-section at 2.76 TeV as comparison, the \Jpsi \raa is measured in the forward muon spectrometer to be about 0.5, with very little centrality dependence. This value is about a factor 2 larger than that measured by PHENIX with muons in the forward region; the difference is less but still significant when comparing with PHENIX electrons at midrapidity. The suppression is also significantly less than that reported by ATLAS. However, the ATLAS results are for high \npt and closer to midrapidity, which would indicate that the suppression either increases with transverse momentum or decreases with rapidity.

While these results are very intriguing, there are still a number of unknowns to be clarified before any firm conclusions can be drawn concerning \Jpsi production at LHC. Besides a small correction for feed-down from B mesons ($\approx 10\%$), the influence of shadowing/saturation effects on nuclear structure functions must be taken into account, as well as 'cold matter' absorption which may (or may not) be important at the LHC energy. For this reason a p--Pb run, which will address these nuclear effects, will be needed and is under discussion for 2013.

Further information on high \npt suppression~\cite{qmharry}, associated particle yields~\cite{qmjanfiete}, heavy flavour~\cite{qmandrea} and \Jpsi production~\cite{qmginez} can be found in the above contributions to these proceedings.

\section{Summary}
At the 2011 Quark Matter conference, the LHC has entered the field of ultra-relativistic heavy ion physics with a very impressive first performance. With a wealth of results barely 6 months after the first collisions, the LHC heavy ion program has benefited from the (too) many years of meticulous R\&D and preparation, a one year 'running in' with proton beams, and, last but not least, a decade of experience and progress made at RHIC. In most results we see a smooth evolution from RHIC to LHC, qualitatively similar but quantitatively different; a sign both that the field is mature enough to be predictable and that the initial conditions (e.g. the energy density) are sufficiently different at both machines so that comparing the results and model predictions will tell us more about the properties of the QGP than looking at either facility alone. With the first low luminosity run, we have just started to explore the 'terra incognita' where the energy reach of LHC makes it unique (e.g. low-x physics, hard processes), and with its very strong and complementary set of detectors we can look forward to many more exciting conferences where quarks (and gluons) matter.

\end{document}